\documentclass[letterpaper]{article} 
\usepackage[]{aaai2027}  
\usepackage[hyphens]{url}  
\usepackage{graphicx} 
\usepackage{natbib}  
\usepackage{caption} 
\usepackage{algorithm}
\usepackage{algorithmic}

\usepackage{newfloat}
\usepackage{listings}
\DeclareCaptionStyle{ruled}{labelfont=normalfont,labelsep=colon,strut=off} 
\floatstyle{ruled}
\newfloat{listing}{tb}{lst}{}
\floatname{listing}{Listing}

\usepackage{booktabs}

\usepackage{xspace}
\newcommand{\ourtool}{\textsc{MindForge}\xspace}

\usepackage[most]{tcolorbox}

\definecolor{MFPromptBlue}{HTML}{2A4B8D}   
\definecolor{MFPromptGreen}{HTML}{1E8449}  
\definecolor{MFPromptRed}{HTML}{B03A2E}    
\definecolor{MFPromptPurple}{HTML}{6C3483} 
\definecolor{MFPromptGray}{HTML}{4D5656}   

\lstdefinestyle{mfprompt}{%
  basicstyle=\ttfamily\footnotesize\color{black},
  breaklines=true,
  breakatwhitespace=false,
  columns=fullflexible,
  keepspaces=true,
  numbers=none,
  xleftmargin=0pt,
  framexleftmargin=0pt,
  aboveskip=0pt,
  belowskip=0pt,
}

\newcommand{\rolepromptfile}[3][MFPromptBlue]{%
  \begin{tcolorbox}[
    enhanced, breakable,
    colframe=#1, colback=#1!5, coltext=black,
    boxrule=0.7pt, leftrule=3pt,
    arc=0pt, outer arc=0pt,
    left=6pt, right=6pt, top=3pt, bottom=3pt,
    before skip=0.7\baselineskip, after skip=0.7\baselineskip,
    coltitle=white, colbacktitle=#1,
    fonttitle=\ttfamily\bfseries\footnotesize,
    title={#2\hfill{\normalfont\textsc{\scriptsize \ourtool agent}}}]
    \lstinputlisting[style=mfprompt]{#3}%
  \end{tcolorbox}%
}
\newcommand{\systempromptfile}[1]{\rolepromptfile[MFPromptBlue]{System}{#1}}
\newcommand{\userpromptfile}[1]{\rolepromptfile[MFPromptGreen]{User}{#1}}

\newcommand{\turnfile}[4][MFPromptBlue]{%
  \begin{tcolorbox}[
    enhanced, breakable,
    colframe=#1, colback=#1!5, coltext=black,
    boxrule=0.7pt, leftrule=3pt,
    arc=0pt, outer arc=0pt,
    left=6pt, right=6pt, top=3pt, bottom=3pt,
    before skip=0.7\baselineskip, after skip=0.7\baselineskip,
    coltitle=white, colbacktitle=#1,
    fonttitle=\ttfamily\bfseries\footnotesize,
    title={#2\hfill{\normalfont\textsc{\scriptsize #3}}}]
    \lstinputlisting[style=mfprompt]{#4}%
  \end{tcolorbox}%
}
\newcommand{\assistantturnfile}[2]{\turnfile[MFPromptPurple]{Assistant}{#1}{#2}}
\newcommand{\toolturnfile}[2]{\turnfile[MFPromptGray]{Tool}{#1}{#2}}

\title{\ourtool: Teaching Small Language Models Whole-Life-Cycle Software Engineering via Source-Free Program Synthesis}
\author{
    Yihao Chen\textsuperscript{\rm 2},
    Shi Chang\textsuperscript{\rm 1},
    Khaled Chawa\textsuperscript{\rm 1},
    Feng Lin\textsuperscript{\rm 1},
    Boyuan Chen\textsuperscript{\rm 1},
    Shaowei Wang\textsuperscript{\rm 3}\corresponding,
    Ahmed E. Hassan\textsuperscript{\rm 2}
}

\affiliations{
    \textsuperscript{\rm 1}Huawei Canada\\
    \textsuperscript{\rm 2}Queen's University\\
    \textsuperscript{\rm 3}University of Manitoba\\
    yihao.chen@queensu.ca,\\
    \{shi.chang, feng.lin, boyuan.chen\}@huawei.com,\\
    khaled.chawa@mail.concordia.ca, shaowei.wang@umanitoba.ca, ahmed@cs.queensu.ca
}
\begin{document}

\makeatletter
\let\originalAAAITitle\@maketitle
\def\@maketitle{%
\originalAAAITitle
\begin{minipage}{\textwidth}
  \centering
  \includegraphics[width=\linewidth]
{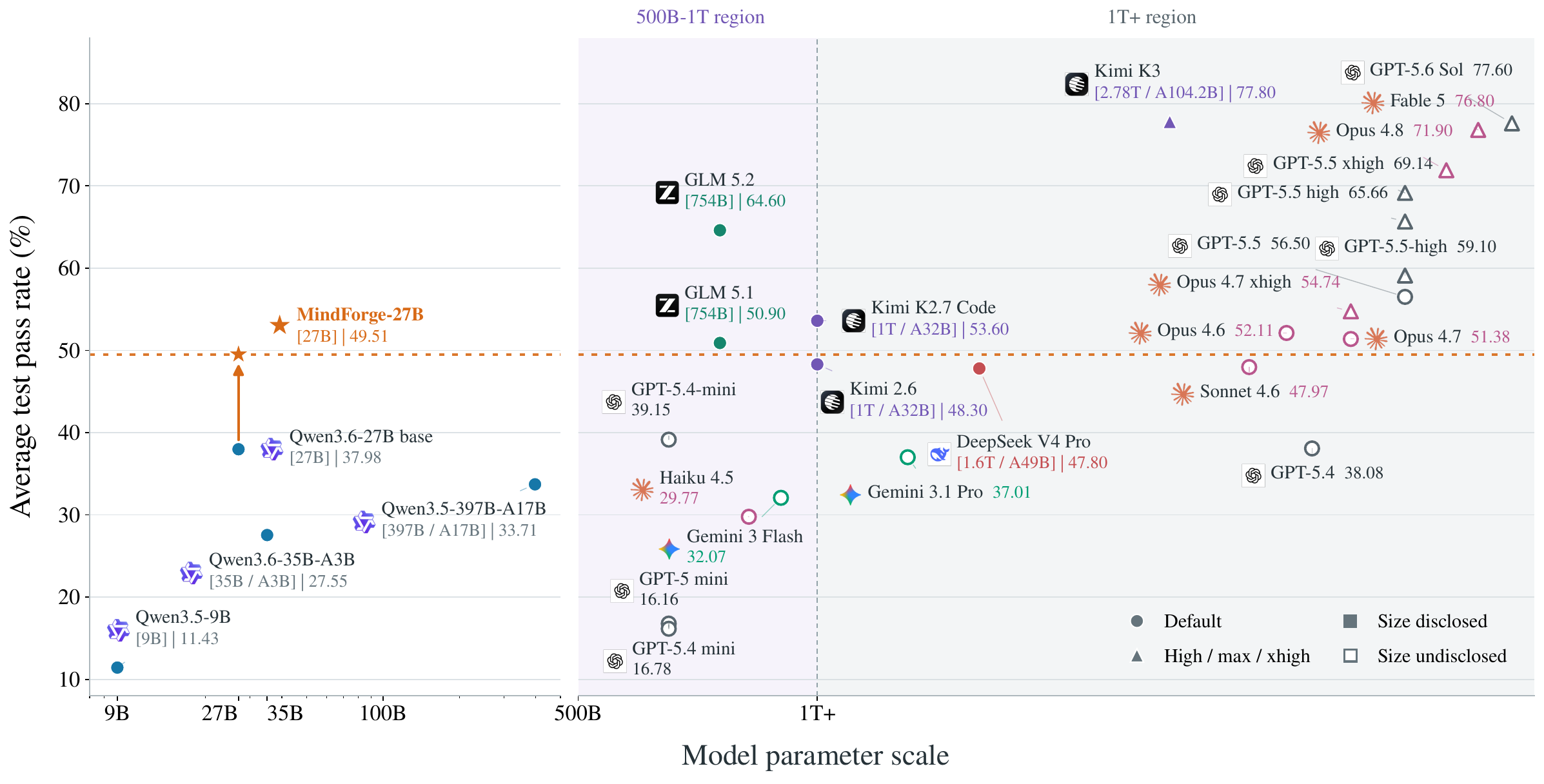}
  \captionof{figure}{
    Average ProgramBench test pass rate by model scale.
    \textbf{\ourtool-27B improves Qwen3.6-27B from 37.98\% to 49.51\%,
    outperforming comparable-size models and achieving performance comparable
    to substantially larger frontier models.} MindForge-27B and the Qwen models
    are evaluated by us on the full 200-instance ProgramBench suite; scores for
    the remaining models are taken from the official ProgramBench leaderboard
    and frontier model technical reports.}
  \label{fig:overall_ranking}
\end{minipage}
}
\makeatother

\maketitle

\begin{abstract}
Coding agents have made substantial progress on software engineering tasks that modify existing codebases, including bug fixing and feature implementation. However, constructing a \emph{complete program from scratch} remains a major challenge: even the frontier models evaluated on ProgramBench fully resolve fewer than 1\% of tasks. One obstacle is the lack of scalable training environments for this from-scratch setting, spanning the whole software engineering life cycle, as existing environment-construction frameworks focus only on a single phase in software development. To address this gap, we introduce \ourtool, an automated pipeline that converts open-source command-line programs into source-free environments that expose only a compiled reference executable and its documentation. Using \ourtool, we construct training environments from repositories disjoint from those in ProgramBench, and curate a high-quality data recipe consisting of program synthesis trajectories using GLM-5.2 as the teacher agent. Fine-tuning Qwen3.6-27B on these trajectories increases its ProgramBench average test pass rate from 37.98\% to 49.51\%, achieving performance comparable to substantially larger frontier models.
Moreover, the fine-tuned model consistently improves over the base model across all seven unseen software engineering benchmarks, spanning long-horizon repository generation and translation, bug fixing, feature implementation, and cross-language issue resolution, with absolute gains of 31.00 points on RepoZero-C2Rust, 14.16 on DeepSWE, 10.70/4.56 on NL2Repo-Bench (with/without tests), 5.04 on SWE-bench Verified, 5.93 on SWE-bench Pro, 5.22 on SWE-bench Multilingual, and 4.94 on FeatBench.

\end{abstract}
\section{Introduction}
\label{sec:introduction}

Coding agents have demonstrated effectiveness across a range of software-engineering tasks, including bug fixing~\citep{jimenez2024swebench}, feature implementation~\citep{zhou2026featurebench,chen2025featbench}, and code completion~\citep{zhang2024codeagent}. These tasks, however, typically require agents to modify or extend an existing codebase. Building a program \emph{entirely from scratch} presents a substantially different and more challenging setting, as the agent must work through the full program-development process: inferring the complete specification from documentation and the observed behavior of the reference program, designing an architecture without an existing implementation to extend, implementing the program, locating and debugging its own bugs, writing tests to expose them, and iteratively refining the program into a working solution. This difficulty is reflected in ProgramBench~\citep{yang2026programbench}, where even frontier models such as GPT-5.5 fully resolve fewer than 1\% of tasks. This result highlights that from-scratch program construction remains an open challenge beyond the scope of existing codebase-modification benchmarks.

A line of work has developed scalable environment-construction pipelines for training coding agents. For example, SWE-Smith synthesizes bug-fixing tasks by introducing faults into existing Python codebases and uses the resulting instances to train coding agents~\citep{yang2025sweSmith}, whereas SWE-Gym pairs real-world software issues with executable repository environments for fine-tuning~\citep{pan2025swegym}. These frameworks have substantially improved agentic performance on bug fixing and feature implementation. However, they assume access to an existing, source-visible codebase and train agents to modify that codebase rather than construct a complete program from scratch. Consequently, scalable training for source-free, end-to-end program construction—and concerns associated with direct source exposure—remain largely unaddressed.

To fill this gap, we introduce the \ourtool pipeline, which combines source-free execution environments with scalable trajectory collection to generate \textit{whole-life-cycle} software engineering trajectories for from-scratch program construction. \ourtool contributes along two axes. First, it converts open-source command-line programs into environments in which an agent is given only a compiled reference executable and its public documentation. The agent must then re-implement the program from scratch by working through the full development process, including specification inference, architecture design, implementation, debugging, testing, and the final submission of a passing build.
This construction mirrors the source-free setting recently used by ProgramBench~\citep{yang2026programbench} and MirrorCode~\citep{adamczewski2026mirrorcode} to \emph{evaluate} long-horizon program re-implementation ability. In contrast, we repurpose this setting to generate \emph{training} data rather than only to evaluate frontier models.

Second, we collect whole-life-cycle program-development trajectories at scale within these environments using a strong teacher agent. We retain only trajectories that produce a valid buildable submission, then refine them to ensure every training example is well formed and free of unresolved errors left unaddressed by the teacher agent. Specifically, we recover trajectories prematurely terminated by transient infrastructure or scaffold failures, and rewrite only the reasoning affected by genuine tool-use mistakes while leaving the teacher agent's tool calls and environment interactions coherent.

We use \ourtool to construct a training dataset of 562 source-free program environments spanning six compiled programming languages. The underlying programs are disjoint from ProgramBench, and the agent has no access to their source code during trajectory collection. Using GLM-5.2 as the teacher agent, we collect 1,001 whole-life-cycle program-development trajectories across these environments. Analysis shows that these trajectories cover multiple key stages of software development, providing rich, multi-stage learning signals: specification exploration, architecture design, bug localization, and refinement appear in 99.1\%, 87.1\%, 59.4\%, and 64.0\% of the trajectories, respectively. We then fine-tune Qwen3.6-27B on these trajectories to reproduce the teacher's end-to-end development process rather than isolated actions. 

Evaluation on ProgramBench shows that fine-tuning increases Qwen3.6-27B's average test pass rate from 37.98\% to 49.51\%. This performance surpasses DeepSeek V4 Pro (47.80\%) and is comparable to substantially larger frontier models, including GLM-5.1 (50.9\%)
and Opus 4.7 (51.38\%). More importantly, the improvement generalizes to seven unseen software-engineering benchmarks and eight
evaluation settings not included in our training recipe, with absolute improvements of 31.00\% on RepoZero C2Rust~\citep{zhang2026repozero},
14.16\% on DeepSWE~\citep{huang2026deepswe}, 10.70\% and 4.56\% on NL2Repo-Bench with and without tests, respectively~\citep{ding2025nl2repo},
5.04\% on SWE-bench Verified~\citep{jimenez2024swebench}, 5.93\% on
SWE-bench Pro~\citep{deng2025swe}, 4.94\% on
FeatBench~\citep{zhou2026featurebench}, and 5.22\% on SWE-bench Multilingual~\citep{zan2026multi}.

Qualitative and quantitative trajectory analysis confirms the gains go
beyond benchmark scores: \ourtool-27B's action distribution moves
substantially closer to its teacher and strong frontier agents, while its command-failure rate stays \emph{lower} despite far longer trajectories -- including one 830-turn, 848-tool-call, 209.5M-token ProgramBench run -- indicating sustained, productive
persistence rather than merely longer, noisier behavior.

We summarize our contributions below:
\begin{itemize}
\item \textbf{A scalable, cross-language pipeline for source-free environment
construction and whole-life-cycle trajectory collection.}
We introduce \ourtool, which automatically converts open-source
command-line programs into reproducible source-free environments. We introduce two refinement procedures to collect high-quality whole-life-cycle program-development trajectories from strong
teacher agents. We release the complete environment-construction,
verification, trajectory-refinement, and distillation pipeline, together
with the resulting environments, trajectories, and \ourtool-27B, to support future research on whole-life-cycle software engineering.

  \item \textbf{A small model matching much larger frontier models.}
  A data and SFT recipe distilling 1,001 whole-life-cycle program synthesis trajectories into a 27B-parameter model
  (Qwen3.6-27B), raising its ProgramBench score from 37.98 to 49.51, moving it into the performance band of substantially larger frontier models, surpassing DeepSeek V4 Pro and nearing Opus 4.7.

  \item \textbf{Generalization beyond the training task.} An empirical analysis showing the same fine-tuned model improves substantially over its base model, and generalizes on \emph{seven} unseen software-engineering benchmarks, spanning out-of-distribution language bug fixing, feature/program implementation, and program translation tasks. 

\end{itemize}

\section{\ourtool: Program Synthesis Trajectories with Coding Agent at Scale}

To collect the trajectories that cover whole life-cycle program synthesis, starting from specification exploration to implementation and bug fixing, \ourtool consists of two phases: first, we build a large pool of reproducible and source-free \emph{executable program environments} from open-source repositories. Second, we collect and refine long-horizon \emph{program synthesis trajectories} inside those environments using a strong teacher agent. We describe the methodology for each phase in sub-sections. The resulting counts and yields are reported separately below.

\subsection{Executable Program Environment Construction}
\label{sec:data-construction-envs}

The goal of this phase is to turn an open-source command-line program into a training environment in which an agent can execute the program without access to its source code. 
We apply the following pipeline to automatically construct an executable environment for each program. A program will be discarded if it fails any of the following steps.

\noindent\textbf{1. Repository selection.} The repository and commit must be fetchable and identifiable, and are pinned so the rest of the pipeline works from a fixed snapshot. We ensure that the source repositories are distinct from those included as benchmark instances to avoid contamination.

\noindent\textbf{2. Offline screening.} An explorer agent reads only the source code and documentation and decides whether the program is a self-contained command-line tool whose behavior can be clearly identified. It does not build, run, or rely on the internet to function (i.e., does not need to access other online services). Candidates that need the public internet, credentials, or special hardware, or whose behavior cannot be clearly identified are rejected here, before any build effort is spent. See Appendix~\ref{sec:exploreAgent}

\noindent\textbf{3. Build discovery.} We design a builder agent to generate a build script that compiles the program directly from source. The script must be self-contained and runnable from a clean repository checkout, so that any later rebuild reflects only what the script itself does, not leftover state from the agent's build session. We only keep programs whose build succeeds under this constraint. We also instruct the agent to record a small set of \emph{behavior checks} (i.e., concrete command invocations with sample inputs), which is used to validate the reproducibility of the build in the next step. See Appendix~\ref{sec:buildAgent}

\noindent\textbf{4. Behavior-equivalence check.}
We verify that the generated build script is deterministic by rebuilding the same pinned source snapshot in a fresh sandbox with no leftover state from the builder. The recorded behavior checks from the previous step are then replayed on both executables (i.e., the independently rebuilt executable and the one produced by the builder), requiring identical exit codes, stdout, and stderr. Any mismatch indicates a non-reproducible environment, in which case the builder retries the instance. This step filters out flaky builds and ensures that the constructed environment is reproducible.

\noindent\textbf{5. Source-free check.} We keep a program environment only if the executable contains no readable form of the program's original source. For the retained instances, we ask the builder agent to compile the reference executable into a native binary (ELF, Mach-O, or PE), and we scan its bytes and strings for markers that would otherwise reveal how the environment was constructed (e.g., if any path in the agent-visible filesystem matches a file in the source snapshot). Any such marker would trigger a retry from the builder agent for re-building. We reject the instance if it eventually still has leakage after re-building.

Each accepted environment produces a Docker image, namely, a \textbf{cleanroom image} which contains only the compiled reference executable and its sanitized public documentation, aligning the format with ProgramBench instances~\cite{yang2026programbench}; this is the only image the trajectory-collection agent ever sees. Since the agent never has source access at any stage, these environments avoid the source-level contamination that affects most repository-derived software-engineering benchmarks.

\subsection{Trajectory Collection and Refinement}
\label{sec:data-construction-trajectories}

Given the constructed environments, we collect complete program synthesis trajectories using \texttt{mini-swe-agent}~\citep{yang2024sweagent} with GLM-5.2~\citep{glm52_2026} as the teacher model, operating directly inside the cleanroom image. The teacher agent is given only the reference executable and its documentation, and must independently elicit a specification, design an architecture, implement it, and iterate to a passing build. We keep only trajectories that the agent ends by issuing the harness's explicit completion command, rather than crashing, timing out, or exhausting its context, and whose emitted \texttt{compile.sh} successfully builds an executable. We do not rely on test suites to rejection sample the trajectories as there are no clear thresholds to define success. This process discards attempts that stalled, looped, or never converged on a buildable solution yet retains diverse synthesis attempts.

After collecting the trajectories, we refine them to ensure that every training example is well-formed and free of unresolved errors left unaddressed by the \emph{teacher agent}, allowing the \emph{student model} to learn from clean supervision rather than spurious signals~\citep{yang2025toolmind,zhou2026offseeker,chen2026signals}. We apply two refinement procedures: (1) \emph{Infrastructure-noise recovery}, which salvages trajectories prematurely terminated by transient infrastructure or scaffold failures before the teacher agent can produce a clean submission, and (2) a lightweight \emph{Reasoning rewrite mechanism}, which repairs trajectories containing genuine tool-use mistakes, by rewriting only the affected reasoning while leaving the remainder of the trajectory unchanged.

\paragraph{1. Infrastructure-noise recovery.} When a teacher trajectory terminates without a clean submission due to transient infrastructure failures (e.g., API errors or service interruptions), we resume execution instead of discarding the trajectory. Specifically, we rewind to the last known healthy step, reconstruct the environment state by replaying all preceding tool calls in a clean environment, and resume execution from that point. This recovery procedure prevents long-horizon trajectories from being abandoned because of infrastructure noise, substantially reducing wasted inference cost.

\paragraph{2. Reasoning rewrite mechanism.} Even a strong teacher agent makes tool-call errors during long autonomous runs, and these errors require more careful handling than simple deletion. We initially detect malformed tool calls and surgically remove them from the trajectory. This removal, however, has a side effect: strong teacher models frequently \emph{reflect} on their own tool-call errors in the reasoning content of later turns, and once the erroring turn is deleted, that reflection no longer refers to anything the model should see, leaving an incoherent non-sequitur in its place. To repair this, instead of direct deletion, we identify follow-up turns whose reasoning plausibly refers to a now-deleted error, and pass each candidate to a repair model (GLM-5.2), which is asked to self-rewrite the orphaned reasoning into a coherent follow-up message, consistent with the trajectory as it now stands. Every proposed rewrite is then screened by a safety check before being accepted, so that a rewrite is applied only when it faithfully repairs the discontinuity, rather than introducing new content. Critically, the rewrite process never touches the tool calls themselves or the environment's recorded responses, so the refinement improves narrative coherence for training purposes without altering the underlying record of what the teacher actually did. See more implementation details and an example in Appendix~\ref{sec:trajectoryRefinement}.

\subsection{Statistics of Collected Programs and Trajectories}
\label{sec:data-construction-stats}

\begin{table}[t]
    \centering
    \small
    \renewcommand{\arraystretch}{1.15}
    \begin{tabular}{@{}l@{\hspace{1.5em}}r@{}}
    \toprule
    \textbf{Item} & \textbf{Size (Mean / Median / Min / Max)} \\
    \midrule
    \# Turns & 181.6\,/\,176\,/\,39\,/\,477 \\
    \# Tokens          & 177K\,/\,182K\,/\,37K\,/\,272K \\
    \bottomrule
    \end{tabular}
    \vspace{-0.1in}
    \caption{Statistics of generated trajectories.}
    \vspace{-0.2in}
    \label{tab:teacher-trajectory-outcomes}
\end{table}

We initially checked out 2,235 candidate repositories from curated ``awesome CLI'' collections on GitHub~\citep{garrettharris2026awesomecli, facchinetti2026awesomeclicsv} and pin each candidate to their latest commit, so that every environment is built from a fixed, reproducible snapshot. After the explorer agent phase, 1,206 programs remain, and among them, 1,002 across 15 languages remain after build discovery and packaging. We then collect teacher trajectories across 562 unique programs spanning six
compiled programming languages -- Go (231, 41.1\%), Rust (212, 37.7\%), C
(87, 15.5\%), C++ (29, 5.2\%), Swift (2, 0.4\%), and TypeScript (1,
0.2\%). 

Running program synthesis on the constructed environments, we collected 1,001 complete trajectories that fit under a fixed rollout budget matching the student model's training token budget (i.e., 256k context window). Table~\ref{tab:teacher-trajectory-outcomes} presents the statistics of the generated trajectories. After the 256K-token training-length filter, 973 of these 1,001 trajectories were used for supervised fine-tuning.

\begin{table}[htbp]
\centering
\begin{tabular}{lcc}
\hline
\textbf{Activity} & \textbf{Coverage} & \textbf{Conditional Coverage} \\ \hline
Spec exploration & 99.1\% & -- \\
Design & 87.1\% & -- \\
Implementation & 99.7\% & -- \\
Bug localization\textsuperscript{*} & 59.4\% & 80.8\% \\
Bug fixing\textsuperscript{*} & 62.6\% & 85.2\% \\
Verification & 83.7\% & -- \\
Refinement\textsuperscript{*} & 64.2\% & 79.5\% \\ \hline
\end{tabular}
\caption{Coverage of each development activity over the 1,001 collected trajectories. \textsuperscript{*}Starred activities need a trigger: localization and fixing require an observed failure (736 trajectories), refinement requires a reliably observed
successful build/test (809). \textbf{Conditional coverage} divides by only those trajectories, since the rest never had the opportunity.}
\label{tab:stage_level_coverage}
\end{table}

To characterize what signal the collected trajectories actually contain, we mine each trajectory for the distinct software-engineering activities it exhibits -- specification exploration, design, implementation, bug localization, bug fixing, verification, and refinement using a rule-based parser. We report each
activity's coverage: the fraction of trajectories in which it appears (see mining details in Appendix~C.1). As shown
in Table~\ref{tab:stage_level_coverage}, the collected trajectories exhibit high coverage across all key stages of software development, confirming that they supply \textit{rich and multi-phase learning signal} rather than repeated exposure to a single activity. Notably, specification exploration and design appear in 99.1\% and 87.1\% of trajectories, respectively -- phases that are absent by construction from narrower, bug-fixing-only training data.
\section{Experimental Settings}\label{sec:exp}

\subsection{Base Model}
\label{sec:exp-base-model}

We fine-tune \texttt{Qwen/Qwen3.6-27B}~\citep{qwen36_27b} as our base student model.
All language-model weights are updated during training; the unused vision components of the checkpoint are frozen. Training and inference are conducted in \texttt{bfloat16} precision.

\subsection{Training Configuration}
\label{sec:exp-training-config}
We fine-tune the model using the MS-Swift~\cite{zhao2024swiftascalablelightweightinfrastructure} framework with its Megatron~\cite{shoeybi2019megatron} backend. We train using sequence packing to minimize padding overhead, with a micro-batch size of 1 and a global batch size of 96 for 8 epochs. Optimization uses AdamW ($\beta_1=0.9$, $\beta_2=0.98$, weight decay $0.04$), a peak learning rate of $4\times10^{-5}$ linearly warmed up over the first 10\% of training steps, decayed to $4\times10^{-6}$ following a cosine schedule, and gradient clipping with a maximum norm of 1.0. We use a random seed of 1105 for both training and data shuffling. The loss is computed only over assistant-generated reasoning, natural language, and tool-call tokens, while system and user messages as well as tool outputs are masked. For evaluation, both the base and fine-tuned models are served under identical settings with a 512K-token context window and reasoning enabled.

\subsection{Evaluation Benchmarks and Metrics}
\label{sec:exp-benchmarks}
\paragraph{Evaluation settings and decontamination.}
We evaluate all eight benchmarks using Mini-SWE-Agent~\cite{yang2024sweagent} as the agent scaffold. During both inference and evaluation, Internet access is disabled to prevent information leakage (e.g., retrieving source code or reference solutions online). We follow each benchmark's official evaluation protocol, reporting its native metric (i.e., resolve rate or pass rate on hidden test sets). Due to the high computational cost of long-horizon benchmarks (ProgramBench, NL2Repo, DeepSWE, and RepoZero), we evaluate each model once on these benchmarks. For the remaining benchmarks, we perform three independent runs and report the mean performance. All reported improvements are statistically significant ($p<0.05$); detailed test results are provided in Appendix~\ref{app:benchmark-significance}. We analyze repository overlap between our training set and all evaluation benchmarks to assess potential data contamination, with particular attention to benchmarks of a similar nature (NL2Repo and RepoZero). We find no repository overlap with five of the seven out-of-distribution benchmarks (RepoZero, SWE-bench Verified, SWE-bench Pro, FeatBench, and NL2Repo). DeepSWE and SWE-bench Multilingual exhibit minimal overlap, involving only two and three repositories, respectively. Nevertheless, these shared repositories correspond to fundamentally different task formulations: our training data consists of end-to-end repository generation without access to the upstream source code, whereas these benchmarks evaluate issue resolution within existing repositories. We therefore consider the risk of evaluation contamination to be negligible. A detailed contamination analysis and repository-overlap statistics are provided in Appendix~\ref{app:repository-overlap-with-ood}.

\paragraph{Primary benchmark.} We evaluate the full \textsc{ProgramBench} suite~\cite{yang2026programbench},
comprising 200 real-world open-source CLI programs (e.g., FFmpeg, SQLite, the PHP interpreter) written in compiled languages -- Rust (107), Go (46), C/C++ (45), Java (1), and Haskell (1). Each instance provides natural-language documentation (README and man page) and an execute-only binary serving as the behavioral oracle for specification elicitation and implementation, and is labeled by difficulty (28 easy, 143 medium, 29 hard)
and reference-binary language. \noindent\emph{\textbf{Average test pass rate (PassRate)}} is our primary metric following previous studies and technical reports~\citep{yang2026programbench,ding2025nl2repo,kimiteam2026kimik3openfrontier}: for each instance $p$, the instance-level pass rate $r_p = k_p / n_p$ is the fraction of hidden test cases the candidate executable passes, and the benchmark-level score $\bar{r}$ is the mean of $r_p$ over all instances. This awards partial credit for implementations that reproduce only a subset of the target's behavior, making it sensitive to incremental gains that a binary resolved/unresolved rate would mask. 

\paragraph{Cross-task generalization benchmarks.} To test whether gains transfer beyond from-scratch program synthesis, we additionally evaluate on seven benchmarks the model was never trained on, spanning distinct software engineering task categories. \emph{Whole-program and repository construction}, closest in spirit to \textsc{ProgramBench} itself, consists of natural-language-to-repository generation (NL2Repo-Bench, 104 tasks)~\citep{ding2025nl2repo} and end-to-end repository translation (RepoZero-C2Rust, 200 tasks)~\citep{zhang2026repozero}. \emph{Issue resolution}, spanning long-horizon development tasks, standard and enterprise-scale repositories, and multiple programming languages, consists of DeepSWE (113 tasks)~\citep{huang2026deepswe}, SWE-bench Verified (500 tasks)~\citep{chowdhury2024swebenchverified}, SWE-bench Pro (731 tasks)~\citep{deng2025swe}, and SWE-bench Multilingual (300 tasks)~\citep{zan2026multi}. \emph{Feature implementation} from natural-language specifications is evaluated using FeatBench (155 tasks)~\citep{chen2025featbench}.

\section{Results}\label{sec:results}

\subsection{Effectiveness on ProgramBench}

\begin{table}[t]
\centering
\begin{tabular}{lccc}
\toprule
\textbf{Model} & \textbf{PassRate (\%)} & \textbf{Win/Lose /Tie} \\
\midrule
GLM-5.2 (teacher) & 64.60   & 36 / 163 / 1 \\
GPT-5.5 & 56.50 & 85 / 114 / 1 \\
Claude Opus 4.7 & 51.38  & 98 / 102 / 0 \\
Sonnet 4.6  & 47.97  & 90 / 110 / 0 \\
GPT-5.4    & 38.08  & 149 / 50 / 1 \\
Qwen3.6-27B (base)   & 37.98          & 152 / 43 / 5 \\
\midrule
\addlinespace[2pt]
\ourtool-27B & 49.51 & -- \\
\bottomrule
\end{tabular}
\caption{The comparison between \ourtool-27B with its base model and other frontier models. Win / Lose / Tie presents the number of instances where MindForge outperforms, underperforms, or ties with each model.}
\vspace{-0.2in}
\label{tab:programbench-core}
\end{table}

\textbf{\ourtool-27B delivers a substantial improvement over its base model on ProgramBench, increasing the average test pass rate from 37.98\% to 49.51\%, an absolute gain of 11.53 points and a 30.4\% relative improvement.} In addition, \ourtool-27B scores 5 instances with over 95\% pass rate, meeting the ``almost resolved'' criterion defined by ProgramBench~\cite{yang2026programbench}, matching the performance of Opus 4.6. \ourtool-27B also scores 100\% on the \texttt{\detokenize{abishekvashok__cmatrix.5c082c6}} instance, a result otherwise achieved only by GPT-5.5 operating in its High and xHigh reasoning modes among the evaluated models from the official leader-board at the time of writing of this paper.

Table~\ref{tab:programbench-core} compares MindForge with its base model. We observe that the gain relative to base model is not driven by a handful of outlier tasks, specifically, \ourtool-27B scores strictly higher on 152 of the 200 tasks (76.0\%), strictly lower on 43 tasks (21.5\%), and ties on the remaining 5 tasks (2.5\%). Given that the base and fine-tuned models share the same architecture and parameter count, this consistent improvement across the majority of tasks indicates that the training recipe is responsible for the gain and provides direct evidence that distilling complete, whole-life-cycle program-synthesis trajectories meaningfully improves a small model's from-scratch software-engineering competence.

\subsection{Cross-Task Generalization}
\label{sec:results-generalization}

\begin{figure} 
\centering 
\includegraphics[width=1\linewidth] {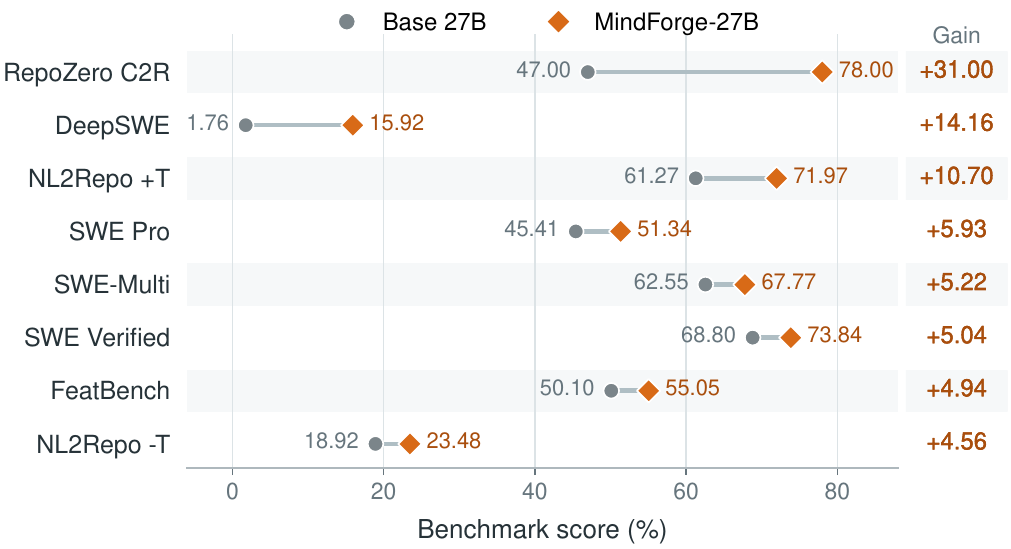} 
\caption{Generalization to seven software-engineering benchmarks across eight evaluation settings. Gain denotes the absolute percentage-point improvement of \ourtool-27B over its 27B base model. None of these benchmarks were used during training.} \vspace{-0.1in}
\label{fig:generalization} 
\end{figure}

\textbf{Fine-tuning on our complete development life cycle trajectories improves the model across diverse software-engineering tasks. More specifically, the fine-tuned model (\ourtool-27B) outperforms its base model across all seven out-of-distribution benchmarks, demonstrating generalization beyond the from-scratch reconstruction setting.} Figure~\ref{fig:generalization} compares \ourtool-27B with its base model across seven benchmarks and eight evaluation settings. The largest absolute gain is on RepoZero C2Rust, where the all-pass rate rises from 47.00\% to 78.00\%, an improvement of 31.00 percentage points (pp). DeepSWE shows the largest relative improvement: its score rises from 1.76\% to 15.92\% (a 9.0$\times$ increase, +14.16 pp). The gains extend to end-to-end repository generation on NL2Repo, both with tests (61.27\% to 71.97\%, +10.70 pp) and without tests (18.92\% to 23.48\%, +4.56 pp). \ourtool-27B also improves on SWE-bench Verified (68.80\% to 73.84\%, +5.04 pp), SWE-bench Pro (45.41\% to 51.34\%, +5.93 pp), FeatBench (50.10\% to 55.05\%, +4.94 pp), and SWE-bench Multilingual (62.55\% to 67.77\%, +5.22 pp).

\subsection{Behavior Analysis}

To understand the behaviors of \ourtool-27B compared with the base model and teacher model, we analyze their trajectories when evaluating ProgramBench. We apply a single command-based classifier, identical across all models, that labels every action in a trajectory as reasoning, inspection, reference probing, implementation editing, building, testing, failure recovery, or submission. Alongside raw operational metrics (i.e., turns, tool-calls, token counts and failure rates), we then report two \emph{transition rates} over consecutive actions: (i) the fraction of reasoning actions that are immediately followed by an implementation edit; and (ii) the fraction of failure-recovery actions that are immediately followed by an implementation edit. Both measure how reliably an agent converts deliberation, or the recovery from a failed command, into an actual change to its code. Table~\ref{tab:Behavior} presents the results of the compared models, and Appendix~\ref{app:behaviorExample} walks through one concrete instance of the failure-recovery transition.

\textbf{Fine-tuning roughly doubles the model's operational engagement with a task while simultaneously lowering its per-command failure rate, showing the extra effort is sustained and productive.}
\ourtool-27B substantially lengthens the model's engagement with a task: mean turns more than double (344.0 to 735.7) and tool calls roughly double (174.4 to 373.0). Total token consumption across the same 200 ProgramBench instances tells a consistent story -- the base model consumed 2.03B tokens in total (mean 10.13M per instance) versus 11.64B for \ourtool-27B (mean 58.22M), a 5.7$\times$ increase -- confirming that end-to-end training encourages longer-horizon reasoning and sustained workflows rather than localized edits, enabling the model to tackle more challenging program-synthesis tasks. This observation complements recent work highlighting the importance of increasing agentic time horizons on complex tasks~\citep{kwa2025measuring,swemarathon_2026}. Meanwhile, the command-failure rate \emph{falls} from 10.98\% to 9.35\% despite this longer horizon, meaning \ourtool-27B sustains more than twice the operational activity of its base model while making proportionally \emph{fewer} mistakes per command. Its tool-call volume now exceeds even the GLM-5.2 teacher's own average (186.6 calls), indicating the fine-tuned model has adopted an even more thorough operational style than its teacher rather than simply copying trajectory length. A coverage analysis confirms this additional activity reflects broader exploration of the reference executable rather than repeated probing, where \ourtool-27B's probing covers 58.39\% of the reference executable against 49.34\% for its base model (more details in Appendix~\ref{app:coverageReport}).

\textbf{Fine-tuning nearly doubles the rate at which the model converts reasoning and failure recovery into actual implementation edits, closing most of the gap to frontier-model procedural discipline.}
The base model edits its implementation after only 27.8\% of reasoning turns and 31.8\% of failure-recovery turns, indicating that reasoning and recovery frequently fail to translate into concrete implementation changes. \ourtool-27B nearly doubles both rates (50.1\% / 48.8\%), moving substantially closer to GLM-5.2 (61.4\% / 64.0\%) and into the same range as GPT-5.4-mini (51.5\% / 37.4\%) and GPT-5.5-high (67.3\% / 70.4\%). Although a gap to the strongest frontier model remains, the trend is clear: fine-tuning on complete whole-life-cycle trajectories teaches the small model to convert a much larger fraction of its reasoning and recovery into concrete implementation changes, narrowing the behavioral gap between the base model and strong frontier agents.

\begin{table*}[htbp]
\centering
\begin{tabular}{l|cccc|cc}
\hline
 & \multicolumn{4}{c}{\textbf{Operational Metrics}} & \multicolumn{2}{c}{\textbf{Behavioral Transitions}} \\ \cline{2-7} 
\textbf{Model} & \textbf{\begin{tabular}[c]{@{}c@{}}Mean\\ Turns\end{tabular}} & \textbf{\begin{tabular}[c]{@{}c@{}}Mean\\ Tool Calls\end{tabular}} & \textbf{\begin{tabular}[c]{@{}c@{}}Mean Peak\\ Prompt Tokens\end{tabular}} & \textbf{\begin{tabular}[c]{@{}c@{}}Command\\ Failure Rate\end{tabular}} & \textbf{\begin{tabular}[c]{@{}c@{}}Editing after\\ Reasoning\end{tabular}} & \textbf{\begin{tabular}[c]{@{}c@{}}Editing after\\ Failure Recovery\end{tabular}} \\ \hline
Qwen3.6-27B (base)     & 344.0      & 174.4 & 115,244    & 10.98\% & 27.8\% & 31.8\% \\
\ourtool-27B & 735.7   & 373.0 & 261,985    & 9.35\%  & 50.1\% & 48.8\% \\
GPT-5.4-mini         & 105.7      & 66.8  & 34,566     & 6.75\%  & 51.5\% & 37.4\% \\
GPT-5.5-high         & 48.6       & 25.0  & 34,558     & 4.51\%  & 67.3\% & 70.4\% \\
GLM-5.2 (teacher) & 369.1 & 186.6 & 212,380 & 7.15\% & 61.4\% & 64.0\% \\ \hline
\end{tabular}
\caption{Agentic Operational Metrics and Behavioral Transitions across Models. We report the mean conversation turns, tool call counts, peak prompt token usage, and command-level failure rates alongside behavioral transition probabilities. Metrics for GLM-5.2 are aggregate statistics over 1,001 teacher trajectories from the training corpus; all other models are evaluated on the same 200 ProgramBench instances. The transition metrics measure the frequency with which an agent actively edits its implementation immediately following a reasoning cycle or a failure recovery event.} 
\vspace{-0.2in}
\label{tab:Behavior}
\end{table*}
\section{Related Work}\label{sec:relatedwork}

\paragraph{Constructing Coding Agent Training Environments.} A separate line of work manufactures verified, non-contaminated environments to train agents, not just evaluate them. SWE-Gym pairs real GitHub issues with executable runtimes and unit tests, and fine-tuning on sampled trajectories yields substantial resolve-rate gains on SWE-bench Verified/Lite, further improved with a trajectory-trained verifier~\citep{pan2025swegym}. SWE-Next scales this further by mining self-verifying pull-request pairs across many repositories~\citep{liang2026swenext}, and R2E-Gym similarly builds procedural environments and hybrid verifiers for open-weight SWE agents~\citep{jain2025r2egym}. SWE-smith instead builds the environment first and generates tasks within it -- procedurally breaking tests in any Python codebase -- to yield a large-scale instance pool~\citep{yang2025sweSmith}. Additionally, SWE-rebench~\citep{badertdinov2025swe}, OpenSWE~\citep{fu2026davinci}, and ScaleSWE~\citep{zhao2026immersion} all scale environment creation to 20,000+ verified environments. While those mentioned studies remain Python-centric, SWE-universe scales this up more than 800,000 environments across 8 languages. All those studies derive tasks from self-contained bugs, issues, or diffs \emph{within} a codebase the agent can see. Beyond language coverage, existing pipelines construct environments around \emph{editing}  visible code for a single phase of engineering work (e.g., bug fixing and feature implementation), while our pipeline constructs environments around \emph{building} code the agent never sees, spanning the whole software engineering development life cycle from specification through a passing build.

\paragraph{Distilling Long Trajectories into Small Models.}Closest to our training methodology is work that trains small, open-weight models by imitating trajectories from stronger models or agents. Lingma-SWE-GPT, SWE-Fixer, SWE-Lego, and Devstral all post-train on staged or filtered issue-resolution trajectories, whether mirroring a developer's process~\citep{ma2024lingmaswegpt}, specializing separate retriever and editor models~\citep{xie2025swefixer}, combining curated real and synthetic issue-resolution trajectories with refined supervised fine-tuning procedures such as error masking and curriculum learning~\citep{tao2026swe}, or iterating by retraining on the model's own rollouts~\citep{rastogi2025devstral}. SWE-Prot\'{e}g\'{e} instead trains a 7B model to selectively call a stronger expert rather than imitate full trajectories~\citep{kon2026sweprotege}. Orthogonal to pure distillation, SWE-RL and CWM apply reinforcement learning -- a rule-based reward over software-evolution data~\citep{wei2025swerl} and large-scale multi-task RL after execution-trace mid-training~\citep{copet2025cwm}, respectively. Our trajectories differ from this literature primarily in \emph{length and life-cycle completeness}: these corpora consist of short-horizon software engineering tasks, with the vast majority (95th percentile) falling within 32K context~\citep{raoof2026openthoughtsagentdatarecipesagentic}, whereas \ourtool trajectories average 181.6 turns and up to 272K tokens each, spanning the full software engineering life cycle—from specification discovery to a passing build—instead of a single localized patch. This whole-life-cycle setting remains under-explored in prior work.

\paragraph{Program and Repository Generation from Scratch.} A recent line of work moves past function-level synthesis to ask whether an LLM (agent) can produce an entire program or repository. RPG frames this as two-stage planning: deciding \emph{what} to build, then \emph{how} to implement it via an explicit repository planning graph~\citep{luo2025rpg}. RepoZero poses generation as \emph{reproduction}: given only API specifications, an agent must re-implement a repository matching a withheld reference implementation, enabling fully automated, execution-based verification~\citep{zhang2026repozero}. DeNovoSWE and NL2Repo-Bench scale this reproduction setting into large training corpora of whole-repository generation tasks to supply long-horizon supervision for training agents rather than only evaluating them~\citep{zhao2026denovoswe, ding2025nl2repo}. These join ProgramBench, which reconstructs complete software projects from only a reference executable and its documentation~\citep{yang2026programbench}, and MirrorCode~\citep{adamczewski2026mirrorcode}, as the emerging cluster of benchmarks treating holistic construction as the unit of evaluation. Different from these prior benchmarks and training corpora, \ourtool contributes an end-to-end pipeline and data recipe that combines automated environment construction, trajectory refinement, and distillation to produce scalable whole-life-cycle software engineering supervision. Together, these components transform executable software into distillation-quality training data for transferring long-horizon software engineering capabilities to small models.

\section{Conclusion}\label{sec:conclusion}
We introduce \ourtool, a pipeline and data recipe that converts open-source command-line programs into source-free software engineering environments, collects whole-life-cycle software engineering trajectories from a strong teacher agent, and refines them into high-quality supervision for distillation. Distilling these trajectories into a small 27B-parameter model raises its ProgramBench score from 37.98\% to 49.51\%, matching substantially larger frontier systems, with gains generalizing to unseen software engineering benchmarks spanning issue resolution, repository generation, translation, and feature implementation. Behavioral analysis confirms these gains reflect genuine transfer of software engineering behavior: the trained model's action distribution converges toward its teacher's while its command-failure rate decreases despite markedly longer trajectories. These results suggest that whole-life-cycle software engineering supervision is an important axis for training capable and efficient software engineering agents.

\bibliography{aaai2027}

\onecolumn
\appendix
\section{Environment Construction Agents Details}
Every agent in the \ourtool pipeline runs on the \texttt{mini-swe-agent}~\citep{yang2024sweagent} harness, driven by the same model, \texttt{Qwen/Qwen3.5-397B-A17B}~\citep{qwen35_2026}. Each agent works inside its own disposable container sandbox, while the pipeline itself is driven from outside that sandbox by the \ourtool orchestrator, which we refer to throughout this appendix as the \emph{host}. The host is ordinary (non-agentic) code: it prepares each sandbox on a Kubernetes cluster, and prepares the inputs for each agentic phase, then collects the agent's artifacts, and finally validates and replays them in a new sandbox, which the agent cannot reach or modify. Each agent produces a structured output, and the host validates it before the instance moves to the next stage. This section describes the three agents. The explorer agent below is the first and cheapest filter, and the other two agents follow the same propose-then-check pattern.

\subsection{Explorer Agent for Initial Screening}\label{sec:exploreAgent}

\paragraph{Role.}
The explorer agent runs the initial screening stage (Section~\ref{sec:data-construction-envs}). For a given repository, it uses \emph{only the source code and documentation} to decide whether the program can become a source-free CLI environment, and it records structured metadata that explains the decision. It runs before any build, so it can reject repositories that cannot become good black-box tasks: tools that need the public internet, credentials, or special hardware to do anything useful; projects with no runnable executable; and programs whose behavior cannot be checked. This happens before the pipeline spends effort building an environment for them. The stage is cheap: it only reads files, and never builds, installs dependencies, runs tests, or uses the network.

\paragraph{Setup.}
The repository is checked out inside a sandbox at a fixed commit, with version-control history removed so the agent cannot read commit messages or upstream references and works only from the file tree. The host also gives the agent a short record of identity and provenance (repository, commit, and source lineage). The agent must treat this record as fixed and cannot change it. The split is intentional: the host sets the identity and provenance facts, and the agent makes every judgment about the program and backs it with evidence. Its only tool is a shell. It inspects the checkout by running bash commands, and is told to keep this inspection short rather than read large lockfiles or test suites in full.

\paragraph{Instructions.}
The \emph{system} prompt sets the role and the main restrictions:

\systempromptfile{LaTeX/prompts/explorer_system.txt}

The user prompt asks the agent to inspect the program and write a single structured judgment. This judgment is a JSON object with a fixed schema, covering: the primary executable and repository kind; dependency, language, and difficulty statistics; license and redistribution status; runtime network needs; observable behavior, determinism, and assertability; documentation quality and oracle-leak risk. For fields that do not affect the decision, the agent writes \texttt{unknown} instead of guessing. Every judgment must cite the repository files it is based on, and those paths must be real files from the checkout; generated build or test outputs do not count as evidence.

The main part of the prompt is a fixed decision policy. The agent must return either \emph{accept} or \emph{reject}; there is no middle ``review'' verdict. Remaining concerns are recorded as review flags on whichever verdict it returns. The decision policy in the \emph{user} prompt is:

\userpromptfile{LaTeX/prompts/explorer_user.txt}

Two parts of this policy are worth noting. First, offline use and public-internet use are two \emph{separate} gates: a program can offer optional online features and still be accepted, as long as its \emph{default} command does useful work on local files, stdin, or a loopback fixture. This is what lets in the many parser, formatter, converter, and local-workflow CLIs that only use the network in optional flags or README examples. Second, being a terminal/TUI program, needing local fixtures, or having weak documentation are review concerns, not blocking failures, because later stages can add fixtures and write clean documentation. Only hard blockers reject a candidate: no executable, a required public service, behavior that cannot be checked, or special hardware.

\paragraph{Expected output.} 
The verdict is expressed as a fixed set of \emph{funnel rules}. Each rule is marked \texttt{pass}, \texttt{fail}, \texttt{not\_applicable}, or \texttt{unknown}, with cited evidence. The rules split into blocking admission gates and non-blocking review signals (Table~\ref{tab:explore-rules}). The link between rules and verdict is fixed and can be checked automatically: an \emph{accept} requires every blocking rule to be \texttt{pass} or \texttt{not\_applicable}; any blocking rule marked \texttt{fail} forces a \emph{reject} and must be listed as a reason; a blocking rule can never be left \texttt{unknown} (missing evidence on a gate counts as a failure); and a review-only rule, such as license or oracle protectability, can never be the sole reason to drop a candidate.

\begin{center}
\small
\captionof{table}{Funnel rules the explorer agent must resolve for each
candidate. Blocking rules are admission gates; review rules are recorded as risk
metadata and never reject a candidate on their own.}
\label{tab:explore-rules}
\begin{tabular}{ll}
\toprule
\textbf{Blocking (admission gates)} & \textbf{Review (risk signals)} \\
\midrule
\texttt{source\_identity\_resolved}            & \texttt{license\_redistributable} \\
\texttt{checkout\_complete}                    & \texttt{first\_party\_behavior\_owned} \\
\texttt{standalone\_executable}                & \texttt{cleanroom\_docs\_assets\_feasible} \\
\texttt{primary\_executable\_selected}         & \texttt{oracle\_protectable} \\
\texttt{runtime\_public\_internet\_not\_required} & \texttt{determinism\_stabilizable} \\
\texttt{offline\_core\_behavior\_viable}       & \texttt{coverage\_instrumentability\_plausible} \\
\texttt{assertability\_viable}                 & \texttt{base\_image\_compatible\_or\_explainable} \\
\bottomrule
\end{tabular}
\end{center}

\paragraph{Verification and keep/discard.} 
Because the judgment decides whether a candidate is admitted or dropped, the host does not accept it as-is. The output is checked twice against the same schema. A validator inside the sandbox gives the agent immediate feedback, and a second validator on the host makes the final decision, so a judgment that was edited inside the sandbox cannot pass itself. The validator checks, among other things: that the JSON is well-formed and has the right schema version; that the agent did not overwrite any host-set identity field; that there is exactly one row per funnel rule with a valid status and the correct blocking/review severity; that the verdict matches the rule outcomes; and that every cited evidence path is a relative, non-generated path that exists in the checkout. This last check stops a judgment from citing files that do not exist. If validation fails, the agent gets the specific errors and is asked to fix only the judgment, without changing a failed gate to pass just to satisfy the checker. The instance is admitted only after a judgment passes both validators.

A candidate is kept only on a validated \emph{accept}. Validated rejects (a failed blocking gate) and judgments that never validate are both dropped, but they are recorded separately, so the dataset-filtering statistics distinguish real screening decisions from schema errors.

\paragraph{Calibration.} 
We calibrated the prompt and validator in two ways before scaling up. On a positive-control set of repositories already known to be good CLI tasks, the policy accepts almost all of them; the few misses were schema errors (invalid evidence or enum values), not wrong decisions. On negative controls, the reject boundaries hold as intended: public-service-only tools, non-executable plugin or library repositories, and programs that need special host hardware or kernel access (for example, a CPU tool that needs a kernel module and model-specific registers, or a backlight tool that needs to write to system device classes) are rejected, while local-file, stdin, loopback, TUI, and local-daemon fixture cases are accepted. Several clauses in the gate policy above were added in response to false rejects found during this calibration: treating README URLs and optional online flags as review flags, allowing wrapper CLIs that call ordinary helper commands, and separating the two network gates.

\subsection{Build Agent}\label{sec:buildAgent}

\paragraph{Role.} 
Once a candidate passes screening, the build agent finds a reproducible way to compile it and produces the two executables the rest of the pipeline needs: the \emph{reference executable} and a \emph{coverage executable} that is compiled with instrumentation. It also writes a small set of host-owned checks (a handful of command invocations with their inputs) that the host later replays to confirm the two executables behave the same and that coverage works. Its job is to make the program build and run, and provide us with a way to verify it does.

\paragraph{Setup.} 
The build agent works in a source-visible sandbox at the pinned commit, with host networking enabled because building often needs to download packages. The base image ships common toolchains (Python with \texttt{uv}, Node with \texttt{nvm}, Bun, Go, Rust with LLVM coverage tools, C/C++ with \texttt{gcc}/\texttt{clang}/\texttt{gcov}/\texttt{make}/\texttt{cmake}, Java with Maven/Gradle, and common development headers), so most repositories build without installing a new toolchain. The agent runs under the standard  \texttt{mini-swe-agent}~\citep{yang2024sweagent} scaffold, capped at 300 steps and a two-hour wall clock.

\paragraph{Instructions.}
The system prompt states the role:

\systempromptfile{LaTeX/prompts/build_system.txt}

The task prompt asks the agent to discover the build and write the required files. Its central rule is that the build must be captured in one self-contained script, \texttt{mindforge\_build.sh}, that runs from a clean checkout: the host re-runs this script in a fresh sandbox, so only steps written into the script are trusted, and anything the agent installs interactively during exploration does not count. The script must build both executables at fixed paths (the reference executable and, when possible, the coverage executable), and the coverage build must instrument the \emph{same} entry point rather than a different code path, so the two executables behave the same. The agent also records, in a JSON file: a \emph{runtime} description of how to launch the reference executable (interpreter or native, version, entry point, environment, whether it needs a terminal); a set of \emph{behavior checks} (command invocations with their inputs); the \emph{coverage tool} to use; and the list of first-party source files its coverage checks are expected to exercise. If the program cannot be built or covered this way, the agent instead reports a give-up with a category and a reason. The full task prompt is:

\userpromptfile{LaTeX/prompts/build_user.txt}

\paragraph{Expected output.} 
The agent fills a fixed template and writes one JSON object (schema \texttt{programbench.build\_discovery.v1}). Its status is either \texttt{ready} or \texttt{give\_up}. A \texttt{ready} record must name the build script and the two executable paths, a valid runtime description, a non-empty set of behavior checks, a coverage tool, and a non-empty list of first-party coverage targets. To keep the output checkable, both the coverage tool and the give-up category are drawn from fixed sets (Table~\ref{tab:build-vocab}); the agent cannot invent a custom coverage tool or parser.

\begin{center}
\small
\captionof{table}{Closed vocabularies in the build agent's output: the supported
host coverage tools, and the categories a give-up must use.}
\label{tab:build-vocab}
\begin{tabular}{ll}
\toprule
\textbf{Supported coverage tools} & \textbf{Give-up categories} \\
\midrule
\texttt{go\_covdata}      & \texttt{coverage\_tool\_not\_supported} \\
\texttt{coverage\_py}     & \texttt{build\_system\_not\_replayable} \\
\texttt{llvm\_cov}        & \texttt{not\_a\_cli} \\
\texttt{gcov}             & \texttt{missing\_required\_dependencies} \\
\texttt{c8\_v8}           & \texttt{binary\_requires\_public\_network} \\
\texttt{bun\_lcov}        & \texttt{host\_equivalence\_not\_constructible} \\
\texttt{shell\_xtrace}    & \texttt{other} \\
\texttt{ruby\_simplecov}  & \\
\texttt{perl\_devel\_cover} & \\
\texttt{jacoco}           & \\
\bottomrule
\end{tabular}
\end{center}

\paragraph{Verification and keep/discard.} 
As with the explorer, the output is validated twice: a validator inside the sandbox rebuilds from the script and runs the checks so the agent gets immediate feedback, and a host validator re-checks the JSON. The decisive step is a fresh host replay: the host re-materializes the source, re-runs the build script in a new sandbox with no leftover state, and rebuilds both executables. It then runs the behavior checks against both executables and requires the exit code, stdout, and stderr to match between them (the equivalence check). The host records the SHA-256 of the agent-built and replay-built binaries for auditing; an exact hash match is not required, as long as the fresh replay builds both binaries and the checks pass. A candidate is kept only when this replay succeeds. The accepted build script, executables, are the input to cleanroom packaging.

\subsection{Coverage Images}\label{sec:coverageAgent}

In addition to the reference executable, we package an instrumented coverage executable together with the corresponding coverage runtime configuration. The build agent records the coverage tool, build commands, and replay configuration required to regenerate coverage traces. These artifacts are verified during the build replay described in Section~\ref{sec:buildAgent} to ensure that the coverage executable can be rebuilt reproducibly.

The resulting coverage image is not used by the whole-life-cycle data-construction pipeline described in this paper. Instead, it is packaged as an auxiliary artifact for future work, enabling downstream studies that explores execution traces or code coverage measurements during long-horizon agent runs.

\section{Trajectory Refinement Details}
\label{sec:trajectoryRefinement}

The two refinement procedures address different failure modes. Infrastructure-noise recovery preserves useful work when execution is interrupted outside the teacher agent's control, whereas reasoning rewrite repairs a local inconsistency caused by a genuine tool-use mistake. In both cases, the objective is to retain as much of the original successful trajectory as possible rather than restarting or rewriting it wholesale.

\subsection{Infrastructure-Noise Recovery}
\label{sec:infrastructureRecovery}

\begin{figure*}[t]
    \centering
    \includegraphics[width=1\linewidth]{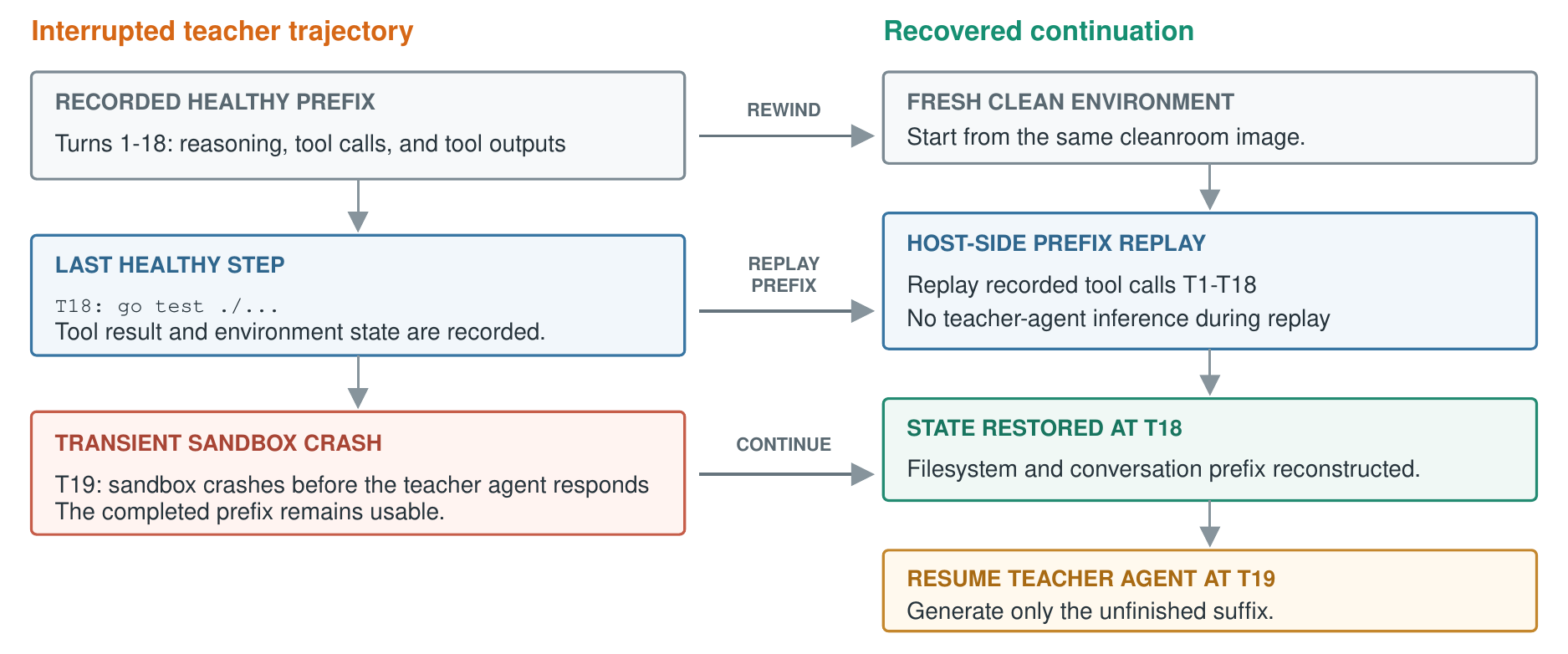}
    \caption{Illustrative infrastructure-noise recovery. After a transient interruption, the host rewinds to the last healthy step, reconstructs state by replaying the recorded tool-call prefix without teacher-agent inference, and resumes generation only for the unfinished suffix.}
    \label{fig:recoveryExp}
\end{figure*}

Recovery is triggered when a transient infrastructure or scaffold failure interrupts an otherwise usable trajectory before the teacher agent can produce a clean submission. We rewind to the last healthy boundary for which both the trajectory record and tool result are complete. The host then creates a fresh copy of the same cleanroom environment and replays the recorded tool calls up to that boundary, in their original order and with their original inputs. Once the filesystem state and conversation prefix have been reconstructed, the teacher agent resumes from the first unfinished turn. Figure~\ref{fig:recoveryExp} illustrates this process.

Replay itself invokes no teacher agent: it re-executes already-recorded tool calls on the host. Recovery therefore avoids regenerating the completed reasoning and action prefix and avoids repeating the intermediate inference calls that produced it. The resumed request still conditions on the reconstructed conversation history, so the saving comes from not solving the completed prefix again, rather than from removing that prefix from the resumed context.

\subsection{Reasoning Rewrite}
\label{sec:rewriteAgent}

\begin{figure*}[t]
    \centering
    \includegraphics[width=1\linewidth]{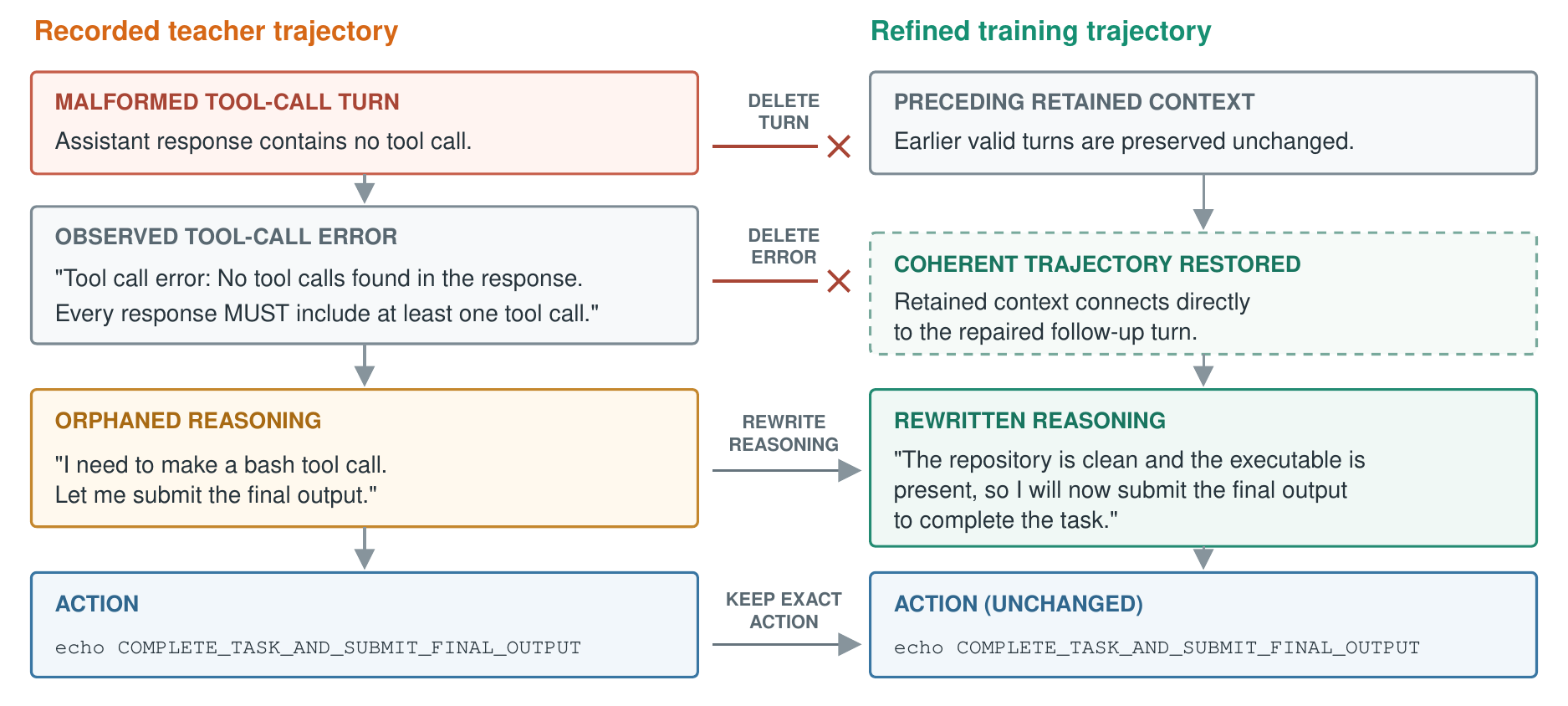}
    \caption{Reasoning rewrite after malformed tool use. Cleanup removes the malformed turn and its scaffold error, the retained context is reconnected to a locally coherent rationale, and the teacher agent's original action is preserved exactly.}
    \label{fig:rewriteExp}
\end{figure*}

A malformed tool-use turn can leave two artifacts: the malformed assistant turn itself and a scaffold-generated error observation. Removing those artifacts can make a later teacher-agent reflection incoherent because it refers to an error that is no longer visible. We therefore identify the affected follow-up turn and ask the repair agent, instantiated with GLM-5.2, to rewrite only its reasoning against the cleaned trajectory context. The repair reconnects the retained prefix to the teacher agent's original next action, as shown in Figure~\ref{fig:rewriteExp}.

The rewrite is deliberately local. The follow-up action, its arguments, subsequent tool calls, and recorded program outputs remain byte-for-byte unchanged; only the affected reasoning text may be replaced. A proposed rewrite is admitted only if the safety check finds it consistent with the visible trajectory and confirms that the preserved structured fields are unchanged. Rejected proposals do not enter the training data and may be regenerated. This constraint lets refinement restore narrative coherence without changing what the teacher agent actually did.

\section{Evaluation Details}

\subsection{Mining various SE activities from trajectories}\label{app:SEActvitiesMining}

\paragraph{Scope and unit of analysis.} We apply a deterministic rule-based parser to the frozen corpus of 1,001 post-rewrite trajectories. A trajectory is the unit of analysis and receives at most one presence flag for each activity, regardless of how many times that activity occurs. The activities are not mutually exclusive: a single trajectory may contribute to several rows of Table~\ref{tab:stage_level_coverage}. The parser processes the recorded events in temporal order, pairs each tool call with its return code, and labels source edits, reference-program probes, inspections, builds, and tests. It then matches the event sequences in Table~\ref{tab:se-activity-rules}. Design is identified from reasoning text because it has no distinctive shell action; the other six corpus-level rates are determined by observable event sequences. Their lexical patterns are used only to retrieve readable examples. We use neither an LLM classifier nor manual annotation to produce the corpus-level counts.

\paragraph{Failure and success rules.} An observed failure is a build or verification event with a final return code from 1 to 127; return code 124 from an expected \texttt{timeout} is excluded. A reliable success requires return code 0 with the check as the final unmasked shell operation. Heredoc bodies are removed before commands are classified, preventing source text inside a file write from being mistaken for an executed command. The windows above are measured in tool events, not assistant turns.

\paragraph{Counting and conditional coverage.} Exploration, design, implementation, and verification use all 1,001 trajectories as their denominator, yielding 992, 872, 998, and 838 matched trajectories, respectively. Localization and fixing are meaningful only when a failure is observed: 736 trajectories contain such an opportunity, of which 595 contain localization and 627 contain a subsequent edit under the rules above. Refinement is conditioned on the 809 trajectories with a reliably observed successful check; 643 continue editing within the specified window. Thus, for an opportunity-dependent activity, conditional coverage is \(N_{\mathrm{matched}}/N_{\mathrm{opportunity}}\), whereas corpus coverage is \(N_{\mathrm{matched}}/1{,}001\). These denominators explain why the two columns in Table~\ref{tab:stage_level_coverage} differ.

\paragraph{Interpretation.} The rules establish that the trajectories expose observable signals for each development stage; they do not recover latent intent. In particular, implicit design can be missed, an intentionally failing negative test can resemble a bug-triggering failure, and a post-success edit is only an operational proxy for refinement. Consequently, this analysis supports the presence of broad, multi-stage development signals but does not by itself establish that those signals caused the downstream generalization gains.

\paragraph{Limitations and future validation.} Rule-based matching can produce both false negatives and false positives: implicit activities may contain none of the selected phrases or command signatures, while an incidental keyword, an expected test failure, or an unrelated edit within a temporal window may satisfy a rule without expressing the intended development activity. The fixed 10- and 15-event windows also trade recall against precision and do not capture every valid development sequence. Future work can validate and calibrate these estimates using a stratified manual inspection or an LLM judge evaluated against human annotations, then report per-stage precision, recall, and uncertainty alongside rule-based coverage. Such semantic validation would refine the prevalence estimates; it would not by itself establish a causal relationship between these activities and downstream generalization.

\begin{table*}[t]
\centering
\small
\caption{Operational rules used to identify software engineering activities. The lexical column lists the complete case-insensitive pattern families used by the parser. Only the design patterns determine a corpus-level rate; patterns for the other activities retrieve examples but do not affect their event-based rates. Brackets denote optional text, slashes denote alternatives, and ``\ldots'' denotes intervening text.}
\label{tab:se-activity-rules}
\begin{tabular}{p{0.13\textwidth}p{0.51\textwidth}p{0.27\textwidth}}
\toprule
\textbf{Activity} & \textbf{Corpus-level presence rule} & \textbf{Reasoning-text patterns} \\
\midrule
Specification discovery / exploration & The supplied reference executable is invoked before the first source edit. Recognized forms are \texttt{./executable}, \texttt{/workspace/executable}, and \texttt{/home/<user>/executable}. In this source-free setting, this probing operationalizes specification discovery from observable behavior rather than elicitation from a user. & \texttt{explor...executable}; \texttt{probe...executable}; \texttt{understand...behavio[r]}; \texttt{test...reference executable} \\
Design & Before the first source edit, assistant reasoning contains an explicit planning expression. This is the only lexical rule used to compute a corpus-level rate. & \texttt{implementation plan}; \texttt{architecture}; \texttt{design}; \texttt{my plan}; \texttt{approach is/will be}; \texttt{strategy} \\
Implementation & A source or build file is modified using \texttt{apply\_patch}, \texttt{sed -i}, redirected \texttt{cat}/\texttt{tee}, a Python file write, or a source formatter. Writes to temporary fixtures, dependency installation, and Git bookkeeping are excluded. & \texttt{now I'll/let's implement}; \texttt{implementing}; \texttt{create the implementation}; \texttt{start implementing} \\
Bug localization & After an observed failed build or test, a targeted inspection, reference probe, or another build/test occurs before a source edit, with that edit no more than 10 tool events after the failure. Targeted inspections include \texttt{rg}/\texttt{grep}, \texttt{cat}/\texttt{head}/\texttt{tail}, \texttt{od}/\texttt{xxd}, \texttt{git diff/show}, \texttt{sed -n}, and \texttt{nl}. & \texttt{root cause}; \texttt{I see/found the issue/bug/problem}; \texttt{the issue/bug/problem is}; \texttt{debugging}; \texttt{diagnos...} \\
Bug fixing & A source edit occurs within 10 tool events after an observed failed build or test. A later successful check is not required for this stage-presence flag. & \texttt{I'll fix}; \texttt{let me fix}; \texttt{need to fix}; \texttt{fix this}; \texttt{apply a/the fix}; \texttt{correct this}; \texttt{I need to move/change/remove/add/update/rewrite/replace/correct} \\
Verification & After a source edit, the trajectory runs \texttt{pytest}/\texttt{unittest}, \texttt{cargo/go test}, \texttt{ctest}, \texttt{make test/check}, \texttt{npm test}, \texttt{diff}/\texttt{cmp}, a static check (\texttt{go vet}, \texttt{cargo clippy}, \texttt{mypy}, or \texttt{shellcheck}), or a custom Python harness that invokes the target program and contains an assertion or equivalent check. This flag records a verification attempt, not necessarily a successful result. & \texttt{let me/I'll/I will [now] run/re-run/execute [the] test(s)/test suite/check(s)}; \texttt{it compiled [successfully]...verify/test}; \texttt{all/the test(s) pass/passed/succeed/match}; \texttt{help/version/output/result(s) [text] [is/are] [byte-]identical/match(es)} \\
Refinement & The trajectory continues editing within 15 tool events after a reliably observed successful build or test. A second successful check is not required for the stage-presence flag. & \texttt{refactor}; \texttt{optimize/improve/enhance/tune the/this}; \texttt{make it/this/the <item> more robust}; \texttt{clean up...dead code}; \texttt{finalize/finalizing/set up...compile.sh/build script} \\
\bottomrule
\end{tabular}
\end{table*}

\subsection{Repository Overlap with Generalization Benchmarks}
\label{app:repository-overlap-with-ood}

We compare the 562 repositories used to construct the ProgramBench training
environments against the repository identities underlying all generalization
benchmarks in Figure~\ref{fig:generalization}. Five repository identities
overlap, covering 17 evaluation instances: three DeepSWE instances and 14
SWE-bench Multilingual instances. We find no repository-identity overlap for
RepoZero C2Rust, SWE-bench Verified, SWE-bench Pro, or FeatBench. NL2Repo's
public metadata exposes target package names rather than GitHub repository
identities; comparing these package names finds no confirmed source-project
overlap, and its two evaluation settings use the same 104 target packages.
Table~\ref{tab:ood-repository-overlap} lists every overlapping instance and
its verifier outcome for the base model and \ourtool-27B. On DeepSWE, both
models fail all three overlapping instances (0/3). On SWE-bench Multilingual,
the base model passes 11/14 and \ourtool-27B passes 12/14; only
\texttt{jqlang\_\_jq-2750} changes from fail to pass.

ProgramBench trains the model to implement programs from scratch using only a reference binary, without seeing the original source code,
whereas DeepSWE evaluates newly authored engineering tasks in existing codebases~\citep{huang2026deepswe}, and SWE-bench Multilingual evaluates issue
resolution in existing codebases~\citep{zan2026multi}. The evaluation prompts, tests, and reference solutions are not used during training. Moreover, the benchmark base commits differ from the commits used to construct the
ProgramBench environments. Therefore, we find that the repository overlap does not imply task or solution leakage.

\begin{table}[t]
\centering
\small
\caption{Verifier outcomes on evaluation instances whose underlying repository
identity also occurs in the 562-repository ProgramBench training pool. Pass
denotes verifier reward 1. The training manifest uses the former repository
name \texttt{stedolan/jq}; GitHub redirects it to \texttt{jqlang/jq}, so they
are treated as the same repository.}
\label{tab:ood-repository-overlap}
\begin{tabular}{lllcc}
\toprule
\textbf{Benchmark} & \textbf{Repository} & \textbf{Instance ID} &
\textbf{Base} & \textbf{\ourtool-27B} \\
\midrule
DeepSWE & \texttt{tomwright/dasel} & \texttt{dasel-html-document-format} & Fail & Fail \\
DeepSWE & \texttt{owloops/updo} & \texttt{updo-policy-alerting} & Fail & Fail \\
DeepSWE & \texttt{go-task/task} & \texttt{task-task-graph-export} & Fail & Fail \\
\textbf{DeepSWE overlap} & & \textbf{Pass rate} & \textbf{0/3} & \textbf{0/3} \\
\midrule
SWE-Multi & \texttt{nushell/nushell} & \texttt{nushell\_\_nushell-12901} & Pass & Pass \\
SWE-Multi & \texttt{nushell/nushell} & \texttt{nushell\_\_nushell-12950} & Pass & Pass \\
SWE-Multi & \texttt{nushell/nushell} & \texttt{nushell\_\_nushell-13246} & Pass & Pass \\
SWE-Multi & \texttt{nushell/nushell} & \texttt{nushell\_\_nushell-13605} & Pass & Pass \\
SWE-Multi & \texttt{nushell/nushell} & \texttt{nushell\_\_nushell-13831} & Pass & Pass \\
SWE-Multi & \texttt{jqlang/jq} & \texttt{jqlang\_\_jq-2235} & Pass & Pass \\
SWE-Multi & \texttt{jqlang/jq} & \texttt{jqlang\_\_jq-2598} & Pass & Pass \\
SWE-Multi & \texttt{jqlang/jq} & \texttt{jqlang\_\_jq-2650} & Fail & Fail \\
SWE-Multi & \texttt{jqlang/jq} & \texttt{jqlang\_\_jq-2658} & Pass & Pass \\
SWE-Multi & \texttt{jqlang/jq} & \texttt{jqlang\_\_jq-2681} & Pass & Pass \\
SWE-Multi & \texttt{jqlang/jq} & \texttt{jqlang\_\_jq-2728} & Fail & Fail \\
SWE-Multi & \texttt{jqlang/jq} & \texttt{jqlang\_\_jq-2750} & Fail & Pass \\
SWE-Multi & \texttt{jqlang/jq} & \texttt{jqlang\_\_jq-2839} & Pass & Pass \\
SWE-Multi & \texttt{jqlang/jq} & \texttt{jqlang\_\_jq-2919} & Pass & Pass \\
\textbf{SWE-Multi overlap} & & \textbf{Pass rate} & \textbf{11/14} & \textbf{12/14} \\
\bottomrule
\end{tabular}
\end{table}

\section{Statistical Significance Analysis}
\label{app:benchmark-significance}
We assess improvements using paired task-level outcomes. Across ProgramBench and
the eight out-of-distribution evaluation settings, every improvement is
statistically significant after Holm correction for nine comparisons
($p_{\mathrm{Holm}}<0.05$). On ProgramBench, a paired Wilcoxon signed-rank test
gives $p=2.38\times10^{-14}$ ($p_{\mathrm{Holm}}=1.90\times10^{-13}$), with a
paired-bootstrap 95\% confidence interval of [8.34, 14.74] percentage points for
the improvement. Among the out-of-distribution evaluations, Holm-adjusted
$p$-values range from $5.75\times10^{-14}$ on RepoZero-C2Rust to $0.033$ on
FeatBench; all paired-bootstrap confidence intervals exclude zero. Full testing
and run-accounting details are provided in Appendix~\ref{sec:statistical-tests}.

\subsection{Statistical Testing and Run Accounting}
\label{sec:statistical-tests}

\paragraph{Experimental units and repeated runs.}
The unit of analysis is a benchmark task, paired between Qwen3.6-27B and
\ourtool-27B. ProgramBench, DeepSWE, both NL2Repo-Bench settings, and RepoZero-C2Rust
were evaluated once due to high inference costs in long-horizon benchmarks. SWE-bench Pro, SWE-bench Verified, SWE-bench Multilingual,
and FeatBench were each evaluated with three runs per model. For a three-run
benchmark, we first average each task's outcome across the three runs and then
compare the resulting paired task means. 

\paragraph{Tests and confidence intervals.}
For a single-run benchmark with paired binary outcomes, we use a two-sided exact
McNemar test. This applies to DeepSWE and RepoZero-C2Rust. We use a two-sided
Wilcoxon signed-rank test for ProgramBench and NL2Repo-Bench, whose task-level scores
can be fractional, and for the four three-run benchmarks, whose per-task means
can take values between zero and one. We report percentile 95\% confidence
intervals for mean score differences using 100,000 paired task-level bootstrap
resamples. To account for the nine comparisons comprising ProgramBench and the
eight out-of-distribution settings, we adjust all $p$-values together using
Holm's step-down procedure.

\begin{table*}[t]
\centering
\small
\setlength{\tabcolsep}{4.2pt}
\caption{Paired significance tests. Gain and confidence intervals are in
percentage points. The run count is per model. Reported $p$-values are two-sided
and adjusted jointly across all nine comparisons using Holm's procedure.}
\label{tab:paired-significance}
\begin{tabular}{lrrrrl}
\toprule
Evaluation setting & Tasks & Runs & Gain [95\% CI] & $p_{\mathrm{Holm}}$ & Test \\
\midrule
ProgramBench & 200 & 1 & $11.53\ [8.34,14.74]$ & $1.90\!\times\!10^{-13}$ & Wilcoxon \\
RepoZero-C2Rust & 200 & 1 & $31.00\ [23.90,38.10]$ & $5.75\!\times\!10^{-14}$ & Exact McNemar \\
DeepSWE & 113 & 1 & $14.16\ [7.08,21.24]$ & $1.61\!\times\!10^{-3}$ & Exact McNemar \\
NL2Repo-Bench (with tests) & 104 & 1 & $10.70\ [4.06,17.42]$ & $4.43\!\times\!10^{-4}$ & Wilcoxon \\
NL2Repo-Bench (without tests) & 104 & 1 & $4.56\ [0.21,8.91]$ & $4.41\!\times\!10^{-3}$ & Wilcoxon \\
SWE-bench Pro & 731 & 3 & $5.93\ [4.01,7.89]$ & $1.68\!\times\!10^{-8}$ & Wilcoxon \\
SWE-bench Verified & 500 & 3 & $4.80\ [2.73,6.93]$ & $2.52\!\times\!10^{-4}$ & Wilcoxon \\
SWE-bench Multilingual & 300 & 3 & $5.33\ [2.11,8.67]$ & $2.33\!\times\!10^{-3}$ & Wilcoxon \\
FeatBench & 155 & 3 & $4.95\ [0.22,9.68]$ & $0.033$ & Wilcoxon \\
\bottomrule
\end{tabular}
\end{table*}

For the binary one-run comparisons, DeepSWE has 18 improved and 2 regressed
tasks (exact McNemar $p=4.02\times10^{-4}$), while RepoZero-C2Rust has 67
improved and 5 regressed tasks (exact McNemar $p=6.39\times10^{-15}$). On
ProgramBench, 152 task scores improve, 43 decrease, and 5 are tied; the paired
Wilcoxon test gives $p=2.38\times10^{-14}$. These unadjusted values are reported
for transparency; the conclusions use the jointly Holm-adjusted values in
Table~\ref{tab:paired-significance}.

\section{Examples}\label{app:examples}

\begin{figure*}
    \centering
    \includegraphics[width=1\linewidth]{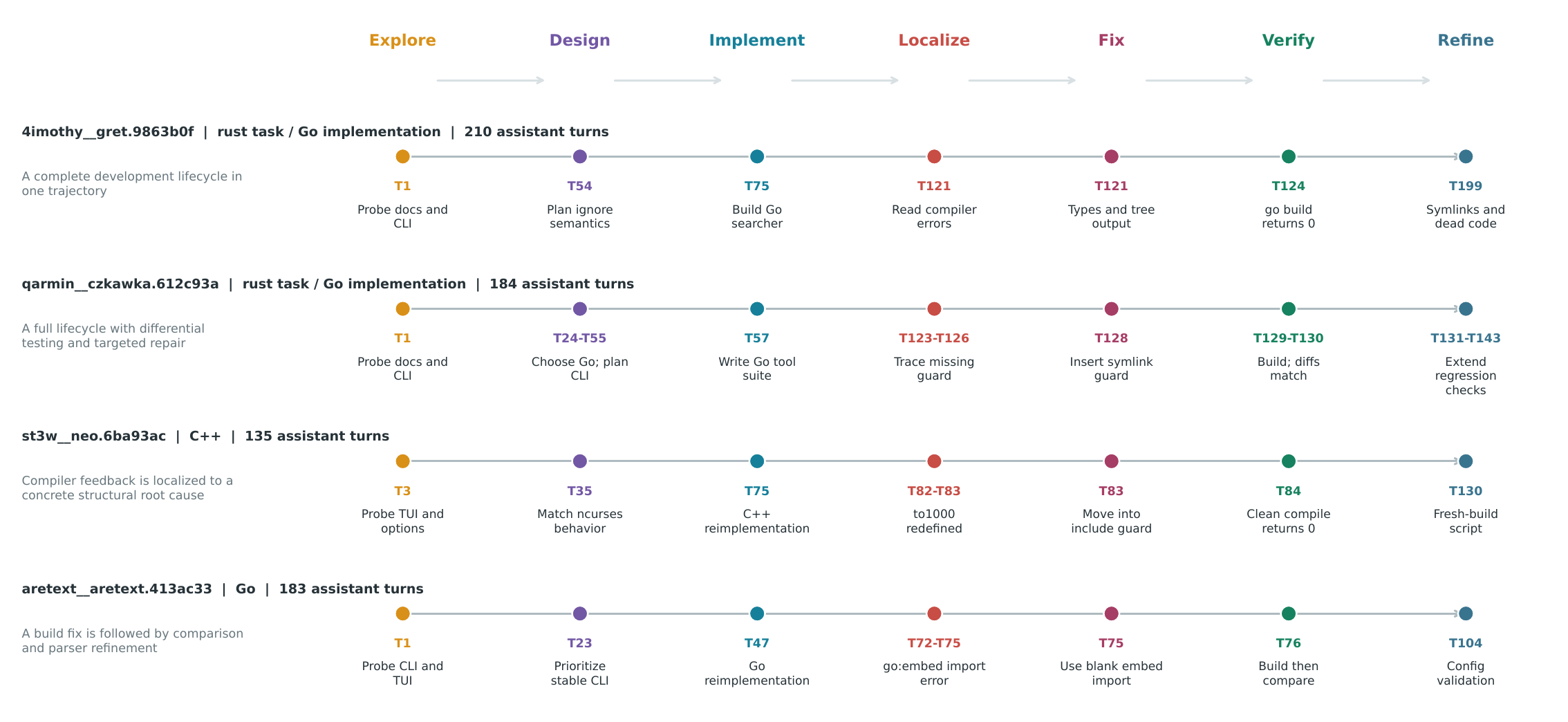}
    \caption{An example of four manually examined trajectories that excercise different software engineering life-cycle activities in various turns.}
    \label{fig:placeholder}
\end{figure*}

\subsection{Examples of editing after failure recovery}\label{app:behaviorExample}
We show what the \emph{editing after failure recovery} transition looks like in practice. In each excerpt below the agent has already implemented a candidate program, has just observed its output diverge from the reference executable, and has correctly identified the cause; the two differ only in what they do next. Long tool output is elided with ``\ldots''.

\paragraph{\ourtool-27B: failure, diagnosis, edit.} At turns 442--443 of 482, the agent diffs its own \texttt{-{}-help} output against the reference and finds that the two formats disagree.

\toolturnfile{\ourtool-27B}{LaTeX/transcripts/mf_tool_fail.txt}

The very next action edits its own source: the reasoning names the cause, and the edit sets exactly the \texttt{clap} flag that produces the reference's two-line help layout.

\assistantturnfile{\ourtool-27B}{LaTeX/transcripts/mf_assistant_edit.txt}

Five actions later the help output matches the reference exactly.

\paragraph{Qwen3.6-27B: failure, diagnosis, no edit.} The agent states the required fix and then does not perform it. On \texttt{\detokenize{mkj__dropbear.75f699b}}, at turns 60--90 of 143, it diffs its own \texttt{-h} usage output against the reference and finds the two disagree.

\toolturnfile{Qwen3.6-27B (base)}{LaTeX/transcripts/base_tool_fail.txt}

Six turns later it has localized the defect to a specific function in its own source and says so explicitly.

\assistantturnfile{Qwen3.6-27B (base)}{LaTeX/transcripts/base_assistant_diag.txt}

The action attached to that intention is not the edit it just described, but another invocation of \texttt{./executable -h}. So are most of the actions that follow: across the \emph{29 consecutive non-edit actions} separating the failure from the eventual fix, 16 re-run \texttt{./executable -h} against the same unchanged binary, twice with byte-identical commands producing byte-identical output. This is re-reading rather than investigating: the reference behavior was already captured in the first diff, and \texttt{print\_usage} is never opened. The agent does not modify its source until turn 90, thirty actions after the failure.

Since both agents correctly identify what is wrong, the main difference is in whether a diagnosis is converted into a change to the code. We stress that a delay is not by itself a defect: re-examining the reference before editing is often the right move, and much of the base agent's post-failure probing elsewhere in the corpus is legitimate investigation. What distinguishes the span above is that it is largely redundant, repeatedly re-reading output the agent had already captured, while the fix it had itself identified went unapplied. This is what the transition rate measures: a failed check followed \emph{within one action} by a modification to the implementation, rather than by another probe, a retry, or an unrelated action. \ourtool-27B does this after 48.8\% of failure-recovery actions, compared with 31.8\% for its base counterpart.

\subsection{Coverage report}\label{app:coverageReport}
\textbf{Fine-tuning substantially increases how much of a reference tool's behavior the model actually exercises before attempting to reproduce it.} To measure how thoroughly an agent probes the binary it is asked to reimplement, we build an instrumented coverage image for each of the 200 official ProgramBench instances, rebuilding every reference tool from its pinned upstream commit with native coverage instrumentation (\texttt{go build -cover} for Go, \texttt{-C instrument-coverage} for Rust, \texttt{gcov} for C/C++, JaCoCo for Java, and HPC for Haskell). We then replay each agent's recorded \texttt{./executable} invocations against that image and report line coverage scoped to the tool's own first-party sources. Averaged over all 200 instances, \ourtool-27B exercises 58.39\% of the reference implementation against 49.34\% for its base model (medians 66.47\% and 52.14\%), an absolute gain of 9.05 points at the mean and 14.33 points at the median. As with the pass-rate results, the effect is broad rather than concentrated: \ourtool-27B attains strictly higher coverage on 154 of the 200 instances and strictly lower on only 11, and the number of instances where the agent exercises at least half of the reference tool rises from 104 to 131. These results provide an independent behavioral perspective on the pass-rate improvements: the fine-tuned model not only produces better implementations, but also performs substantially more comprehensive empirical exploration of the target artifact before attempting to reproduce it.

\end{document}